\documentstyle[11pt,newpasp,twoside,epsf]{article}
\def\Halpha{{\,\rm H_{\alpha}}}
\def\etal{{et al.~}}

\def\edcomment#1{\iffalse\marginpar{\raggedright\sl#1\/}\else\relax\fi}
\reversemarginpar

\begin{document}
\title{HST observations of NGC 6240}
\author{Joris Gerssen, Roeland P. van der Marel}
\affil{Space Telescope Science Institute, Baltimore, USA} 
\author{David Axon} 
\affil{University of Hertfordshire, Hatfield, UK}
\author{Chris Mihos} 
\affil{Case Western Reserve University, Cleveland, OH, USA}
\author{Lars Hernquist} 
\affil{Harvard University, Cambridge, MA, USA}
\author{Joshua E. Barnes} 
\affil{Institute for Astronomy, Honolulu, HI, USA}

\begin{abstract}
WFPC2 images and STIS spectroscopic observations are presented of the
double nucleus in the merger system NGC~6240. We find that: (a) the
kinematics of the ionized gas is similar to that of the molecular gas,
despite a different morphology; (b) the gaseous and stellar kinematics
are quite different, suggesting an early merger stage; (c) neither the
gaseous nor the stellar kinematics show an obvious sign of the
supermassive black hole believed to be responsible for the X-ray
emission of NGC 6240; and (d) the steep off-nuclear velocity gradient
is not due to a $\sim 10^{11} M_{\odot}$ black hole, in contrast to
earlier suggestions.
\end{abstract}

\section{Introduction}

The luminous IR galaxy NGC 6240 has both the tidal tails and a double
nucleus that are characteristic of a merging system.  Since its
identification by the IRAS satellite as one the most nearby ULIRGs,
NGC~6240 has been the subject of numerous studies at virtually all
wavelengths.

One of the key results emerging from these studies is that NGC 6240
harbors both a starburst and an AGN. Several lines of evidence
indicate the presence of young stars, including e.g.~the observed
strength of the CO bandhead (Rieke \etal 1985). However, observations
with various X-ray satellites show a strong X-ray component that can
only originate from an AGN (e.g., Vignati \etal 1999). Spectroscopic
line diagnostics in the mid-IR support the view of a composite
AGN/starburst source, but indicate that most of the IR luminosity of
$\sim~5~\times~10^{11}~L_\odot$ must be powered by the starburst
(Genzel \etal 1998).

To improve our understanding of the double nucleus we have studied
this region at high resolution using the Hubble Space Telescope (HST;
GO Programs 6430 and 8261; PI: van der Marel).  We have obtained WFPC
imaging and longslit STIS spectra to determine both the gaseous and
the stellar kinematics.  An additional set of emission line spectra
was obtained some 12 arcsec from the double nucleus to examine the
nature of a large velocity gradient that exists at this location
(Bland-Hawthorn, Wilson \& Tully 1991).

\begin{figure}[t]
\plotone{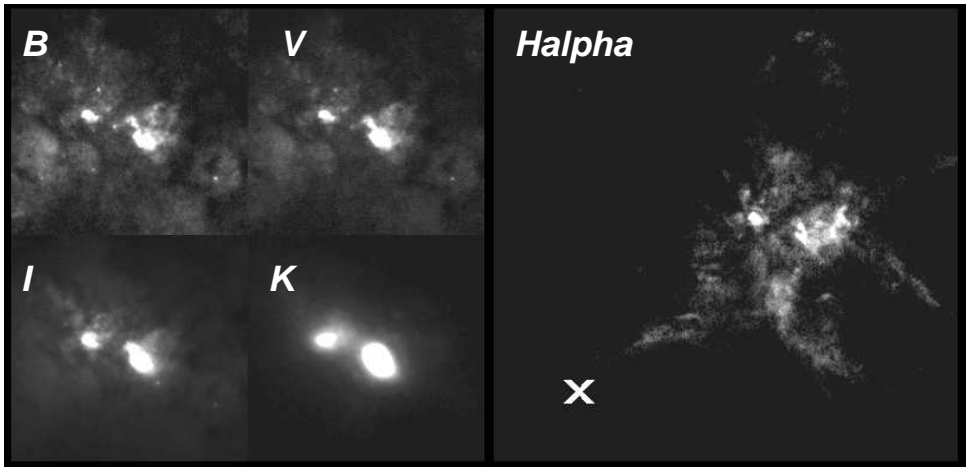}
\caption{{\it (a; left)} Mosaic of HST images in the $B$, $V$, $I$ and
$K$ bands from WFPC2 and NICMOS (each frame is 8 by 8 arcsec). {\it
(b; right)} Narrow-band WFPC2 $\Halpha$+[NII] image (20 by 20 arcsec).
The cross marks the position of the hypothesized black hole that is
discussed in Section~3. The orientation is the same in all panels,
with north in the top left corner.}
\end{figure}

\section{Double Nucleus}

\subsection{Imaging}

The two nuclei of NGC 6240 are separated by 1.8 arcsec or about
0.8~kpc. Figure~1a shows close-ups of our WFPC2 observations in the
$B$, $V$ and $I$ bands (filters F450W, F547M and F814W, respectively),
together with a K-band NICMOS image retrieved from the HST Data
Archive (filter F222M, GO program 7219, PI: Scoville). There is a
clear change in the morphology of the double nucleus as the wavelength
increases, due to strong dust absorption, especially on the southern
nucleus. This is consistent with the previous finding from
lower-resolution ground-based data that the distance between the
flux-centroids of the two nuclei appears to decrease with wavelength
(Schulz \etal 1993).

Figure~1b shows our narrow band WFPC2 $\Halpha$+[NII] image. The
double nucleus is visible also in ionized gas. The large-scale
morphology of the gas has a highly filamentary structure that is
characteristic of a starburst wind (Heckman, Armus \& Miley 1987).

\subsection{Gas Kinematics}

To map the velocity field of the ionized gas in the nuclear region we
obtained a set of STIS $\Halpha$+[NII] emission line spectra with the
slit parallel to the line that connects the nuclei. We used slit
widths of $0.1''$ and $0.2''$ to cover an area with a full width of
$1.1''$. The spectra were analysed by fitting Gaussians to the
emission lines. The best-fitting Gaussian parameters yield the flux,
radial velocity and velocity dispersion. The results are shown as
contour maps as a function of position in Figure~2a--c.

The flux distribution corresponds well with that in the
$\Halpha$+[NII] narrow band image. Although the northern nucleus is
less luminous than the southern nucleus in continuum emission, it is
more centrally concentrated and has a higher peak intensity in the
emission lines.

The velocity field does not show clear signs of rotation around either
of the two nuclei. However, it does show a steep velocity gradient of
400 km s$^{-1}$ arcsec$^{-1}$ between the two nuclei, as shown in
Figure~2d. The velocity field of molecular CO gas obtained by Tacconi
\etal (1999) yields a similar picture, with the same velocity gradient
that we observe in $\Halpha$+[NII] (same peak to peak velocity
amplitude and turn-over radius). While Tacconi \etal observe that the
CO velocity field is highly complex, they are able, with some
simplifying assumptions, to fit it with a model of a rotating disk
between the two nuclei. This is not unreasonable, given that the
observed CO flux peaks between the two nuclei. However, the ionized
gas emission that we observe clearly does not peak between the two
nuclei, cf.~Figure 2a. It is therefore surprising that the ionized gas
and the molecular gas do have very similar kinematics.

Current N-body simulations do not yet have sufficient resolution to
follow the gaseous and stellar kinematics inside the central kpc of
mergers in much detail. However, they do indicate that the dynamics
can be complex and out of equilibrium, so the interpretation of the
gas velocities in NGC 6240 as due to organized disk rotation should be
viewed with some caution. Note that gas disks are often observed
around the nuclei of ULIRGs, but not generally in between the
nuclei.\looseness=-2

\begin{figure}[t]
\plotone{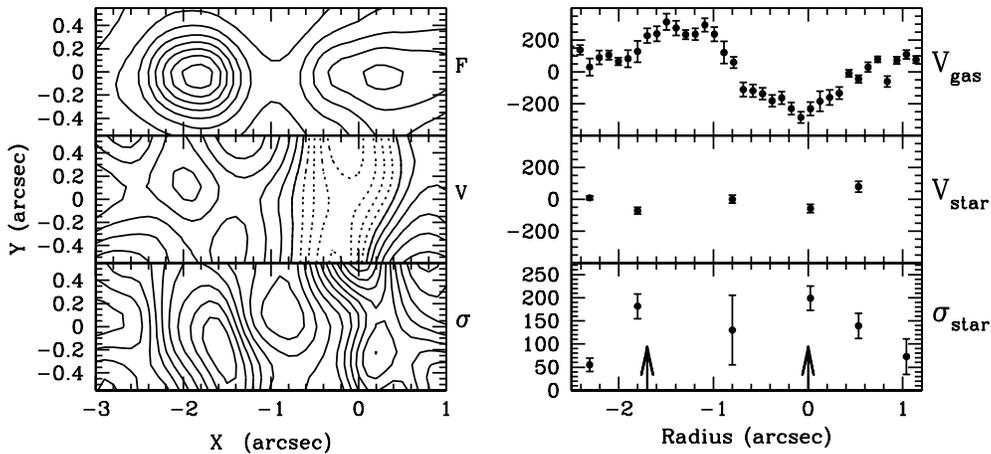}
\caption{Kinematics of the central region of NGC 6240 derived from HST
STIS spectroscopy. The panels on the left are two dimensional contour
plots. The horizontal direction corresponds to the position angle that
connects the nuclei. The northern nucleus is on the left. The panels
on the right are kinematical profiles along the line that connects the
nuclei. {\it (a; top left)} contour plot of the flux distribution of
$\Halpha$+[NII]. The double nucleus is clearly visible. {\it (b;
middle left)} velocity field of the ionized gas; dotted contours
indicate negative velocities. {\it (c; bottom left)} contour map of
the velocity dispersion of the ionized gas. The dispersion peaks near
the two nuclei, and is lower in between. {\it (d; top right)} mean
line-of-sight velocity of the gas along the line that connects the
nuclei. There is no obvious sign of rotation around either of the two
nuclei, but there is a strong velocity gradient between the two
nuclei. {\it (e; middle right)} mean line-of-sight velocity of the
stars. {\it (f; bottom right)} the stellar velocity dispersion. Arrows
indicate the positions of the nuclei.}
\end{figure}

The velocity dispersion map of the ionized gas shows peaks that
roughly coincide with the two nuclei. However, the difference between
the maximum and minimum velocity dispersions is small, less than 25
percent. To first order, the velocity dispersion of the ionized gas is
consistent with $\sim$ 250~km~s$^{-1}$ over the entire nuclear region.
One of the goals of the gas kinematical observations was to find
evidence for broad emission lines that could identify the AGN
responsible for the X-ray emission. However, we see no sign of broad
emission lines or strongly increasing line widths towards either of
the nuclei.

\subsection{Stellar Kinematics}

We also obtained STIS spectra at the near-IR Ca triplet to study the
stellar kinematics in the region of the double nucleus. Due to the
lower S/N of absorption lines it was not possible to make a
two-dimensional map of the kinematics, so instead only a single
spectrum was obtained with the slit along the line connecting the
nuclei. The inferred mean velocities and velocity dispersions of the
stars are shown in Figures~2e,f.

The stellar kinematics differ considerably from the gas kinematics.
The large velocity gradient seen in the ionized gas is absent in the
stellar kinematics and there appears to be very little rotation around
the nuclei. However, a ground based stellar velocity field derived
from the CO bandhead (Tezca \etal 2000) does show rotation around both
nuclei but at large angles to the position angle of the line that
connects the nuclei.

The stellar velocity dispersion (derived after binning rows along the
spectrum to increase signal-to-noise) peaks at the nuclei. At both
nuclei the measured velocity dispersion is around 200 km s$^{-1}$,
well below earlier, spatially unresolved, observations of $\sim$ 350
km s$^{-1}$ (Lester \& Gaffney 1994; Doyon \etal 1994), but consistent
with the more recent observations by Tezca et al. The signal-to-noise
ratio between the two nuclei is too low, even after rebinning, to
reliably determine the stellar velocity dispersion. Tezca \etal find
that in their data the velocity dispersion actually peaks between the
nuclei. However, the dynamical importance of this observation is
unclear because N-body simulations show that projection effects can
lead to spurious peaks in the observed velocity dispersions (Mihos
2000).

Stars are collisionless objects, unlike gas which can be influenced by
shocks, infall and starburst winds. In general stars therefore trace
the gravitational potential of a galaxy more reliably than gas. One of
the goals of the stellar kinematical observations was to find evidence
for an increasing velocity dispersion towards either of the nuclei,
which would have provided evidence for the gravitational influence of
the black hole that is responsible for the X-ray emission. However, we
do not observe the large velocity dispersions that would
unambiguously identify a black hole.

\begin{figure}[t]
\plotone{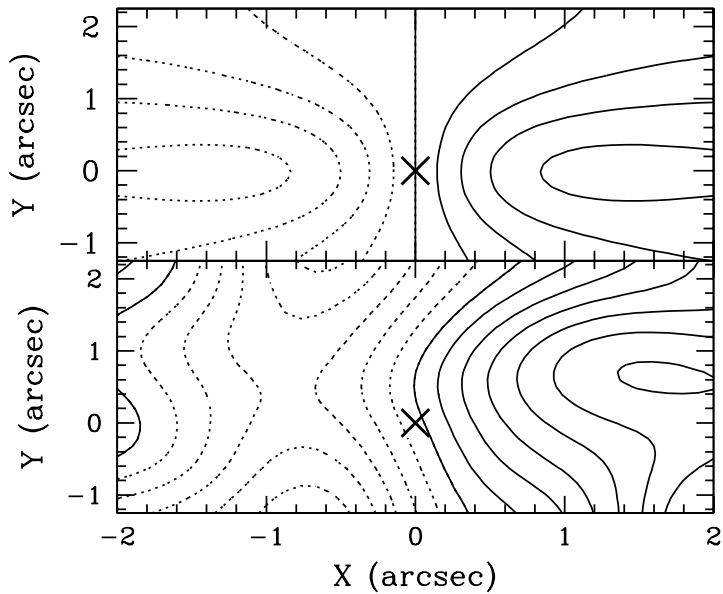}
\caption{The ionized gas velocity field at the position of the
hypothesized off-nuclear black hole (cross; see also Figure~1b). {\it
(a; top)} Model velocity field for a Keplerian disk around a $10^{10}
M_{\odot}$ black hole. {\it (b; bottom)} Velocity field observed with
HST at $\sim 0.5''$ resolution (the model in the top panel was
smoothed to the same resolution). The observed kinematics are not
consistent with the hypothesized black hole. The x-axis of both panels lies
in the direction of position angle 155$^\circ$.}
\label{bhres}
\end{figure}

\section{The Hypothesized Off-Nuclear Black Hole}

Bland-Hawthorn \etal (1991) reported the presence of a steep velocity
gradient in a ground-based $\Halpha$ velocity field of NGC 6240 at a
projected distance of 6 kpc from the double nucleus (marked by a cross
in Figure~1b). This was interpreted with a model in which the velocity
gradient is due to rotation around a black hole of $\sim 10^{11}
M_{\odot}$. This would be quite remarkable, given that the most
massive black hole ever convincingly detected is only $\sim 3 \times
10^9 M_{\odot}$ (e.g., Kormendy \& Gebhardt 2001).  Also, no
counterpart is seen at the position of the hypothesized black hole in
optical continuum, ionized gas, radio emission or X-ray wavelengths.

We have mapped the $\Halpha$ velocity field around the hypothesized
black hole using STIS spectra obtained with the $\sim 0.5''$ wide slit
placed at several parallel positions (Figure~3b). While the resulting
spatial resolution is admittedly low for HST standards, it is still a
few times better than for the best available ground-based
data. Although a steep gradient is observed in the HST data, the
gradient is not steeper than the measured ground based value. This is
not consistent with expectation if there were indeed a $\sim 10^{11}
M_{\odot}$ black hole present. Then the gradient should be steeper when
observed at higher spatial resolution. If the velocity gradient
observed with HST were interpreted with a model of gas orbiting a
black hole, the implied black holes mass would only be $\sim 10^{10}
M_{\odot}$. However, such a model is inconsistent with the observed
two-dimensional structure of the velocity field. The predicted
velocity field at the observational resolution for an inclined
Keplerian disk is shown in Figure~3a. The velocity gradient quickly
flattens perpendicular to the kinematical major axis, by contrast to
the observed gradient. Hence, the observed velocity field is
inconsistent with the signature expected from a black hole. The
location of the observed velocity gradient along the extension of a
filament (see Figure~1b) suggests instead that it is due to kinematic
gradients in the starburst wind, or possibly projection effects.
\looseness=-2

\end{document}